\documentclass{article}
\def\be{\begin{equation}}
\def\ee{\end{equation}}
\def\bea{\begin{eqnarray}}
\def\eea{\end{eqnarray}}
\def\ben{\begin{eqnarray*}}
\def\een{\end{eqnarray*}}
\def\pl{\partial}

\def\btb{\begin{tabular}}
\def\etb{\end{tabular}}
\usepackage{graphicx}

\begin{document}

\begin{center} {\bf\large{On Properties of Vacuum Axial Symmetric Spacetime \\ of
Gravitomagnetic Monopole \\ in Cylindrical Coordinates}}

\vspace*{1cm}

{\it Valeria Kagramanova$^{1,2}$ and  Bobomurat Ahmedov$^{1,2,3}$}

\end{center}

\vspace*{1cm}

\begin{flushleft}

{$^1$Institute of Nuclear Physics, Ulughbek, Tashkent 702132,
Uzbekistan
\\ $^2$Ulugh Begh
Astronomical Institute, Astronomicheskaya 33, Tashkent 700052,
Uzbekistan\\ $^3$International Centre for Theoretical Physics,
Strada Costiera 11, 34014 Trieste, Italy}

\end{flushleft}

\vspace*{1cm}

\begin{abstract}
We investigate general relativistic effects associated with the
gravitomagnetic monopole moment of gravitational source through
the analysis of the motion of test particles and electromagnetic
fields distribution in the spacetime around nonrotating
cylindrical NUT source. We consider the circular motion of test
particles in NUT spacetime, their characteristics and the
dependence of effective potential on the radial coordinate for the
different values of NUT parameter and orbital momentum of test
particles. It is shown that the bounds of stability for circular
orbits are displaced toward the event horizon with the growth of
monopole moment of the NUT object. In addition, we obtain exact
analytical solutions of Maxwell equations for magnetized and
charged cylindrical NUT stars.
\end{abstract}


\vspace*{0.5cm}


\vspace{0.5cm}

\newpage

\section{Introduction}

At present there is no any observational evidence for the
existence of gravitomagnetic monopole or so-called NUT charge
which may be source of gravitomagnetic field in static case.
However there are several attempts and proposals to detect it
through astronomical observations in galactic, extragalactic range
and in Solar system. For example, the anomalous acceleration of
Pioneer satellites~\cite{pioneer} is attempted to explain through
the effect of gravitational field of {\it magnetic mass} on motion
of test particles~\cite{pioneer_nut1,pioneer_nut2}. In recent
papers ~\cite{rahvar,rahnour}, the theoretical effect of magnetic
mass in NUT space on the microlensing light curve and the
possibility of magnetic mass detection using the gravitational
microlensing technique have been studied.

It is not less interesting to study vacuum axial-symmetric
solutions of Einstein equations describing the gravitational field
of nonrotating NUT source. Nowadays cylindrically symmetric
stationary metrics are analyzed due to the fact that they can be
used as simple first order approximation model for describing some
astrophysical objects.  Here we will concentrate in studying
properties of NUT sources related to the cylindrical symmetry. The
NUT metric components generated by the solution to the Newtonian
Laplacian equation corresponding to a Newtonian potential of a rod
of length $2A$ and mass per unit length being equal to ${1}/{2}$
was presented in~\cite{gauthoff}. In the paper~\cite{zonoz97} the
metric of cylindrical NUT source in the approximation of weak
field has been obtained and studied. In the paper~\cite{avas-bobo}
we have shown the metric of Nouri-Zonoz describing spacetime
outside a line gravitomagnetic field could be obtained from
Papapetrou solution of vacuum equations of gravitational field.

 Physical interpretation of NUT
solution was also investigated in the literature. In particular,
 it was shown in recent paper~\cite{manko} that the sources of the NUT
spacetime are two semi-infinite counter-rotating rods of negative
masses and a finite static rod of positive mass. It was also noted
that this model is the only possibility for the NUT solution to be
stationary and possess zero total angular momentum. Although the
metric is not, from the mathematical point of view, cylindrically
symmetric, it is physically cylindrically symmetric, which is
clear from the gravitoelectric and gravitomagnetic fields which
are both explicitly cylindrically symmetric functions.

In our preceding paper~\cite{bobo} the electromagnetic fields of
conducting shell embedded  in the weak gravitational field of
cylindrical gravitomagnetic monopole were investigated and
analytical solutions to the Maxwell equations in this spacetime
were found. Here we extend them to the space time of cylindrical
NUT source obtained in~\cite{manko} as exact solution of Einstein
equations without making any assumption on the strength of
gravitational field. We are interested to study the motion of test
particles and electromagnetic fields in NUT space with the aim to
get tools for studying new important general relativistic effects
which are associated with nondiagonal components of the metric
tensor and have no Newtonian analogues due to their origin.

The paper is organized as follows. In section~\ref{motion} we
study the motion of test particles in spacetime of cylindrical NUT
source. We investigate the effective potential and consider the
influence of NUT parameter on characteristics of particles motion.
In section~\ref{meq} we solve Maxwell equations in the exterior
metric of cylindrical NUT source around (i) magnetized star with
the dipolar magnetic field and (ii) electric charge put on
cylindrical gravitomagnetic monopole. We summarize our results in
section~\ref{conclusion}.

Throughout, we use the units $G=c=1$ and the metric signature
$(-,+,+,+)$. The round and square brackets on the indices denote
symmetrization and antisymmetrization respectively, e.g.,
$A_{(\mu\nu)}=\left(A_{\mu\nu}+A_{\nu\mu}\right)/2$ and
$A_{[\mu\nu]}=\left(A_{\mu\nu}-A_{\nu\mu}\right)/2$.

\section{Particles Motion in a Space of Nonrotating Cylindrical NUT Source}
\label{motion}

The general stationary axially symmetric
solution~\cite{gauthoff,manko} to the vacuum Einstein equations in
the standard Weyl-Papapetrou cylindrical coordinates
$\{t,\rho,\varphi,z\}$ is given by
\be \label{gen_cylind} ds^2 =
f^{-1}\left[e^{2\gamma}(d\rho^2+dz^2) + \rho^2d\varphi^2\right] -
    f\left(dt-\omega d\varphi    \right)^2\ ,
\ee
where the metric coefficients $f,\gamma,\omega$ are functions of
$\rho$ and $z$ only. The explicit expressions for $f,\gamma$, and
$\omega$ have been obtained by Gautreau and Hoffman
~\cite{gauthoff} in the form
\bea \label{gaho} &&f=\frac{A^2(x^2-1)}{(A x+M)^2+l^2}\ , \quad
e^{2\gamma}=\frac{x^2-1}{x^2-y^2}\ , \quad \omega=2ly+C\ ,
\nonumber\\ &&x=\frac{p_{+}+p_{-}}{2A} \ , \quad
y=\frac{p_{+}-p_{-}}{2A} \ , \nonumber\\
 &&p_{\pm}=\sqrt{\rho^2 +(z \pm A)^2} \ , \quad A =
\sqrt{M^2+l^2} \ ,
 \eea
where $C$ is an arbitrary real constant.

Following to~\cite{manko} we will take here $C=0$ (case $C=0$ also
corresponds to the general NUT metric
\begin{equation}
ds^2 =-\alpha[dt-2l\cos\theta d\phi]^2
+\sin^2\theta(r^2+l^2)d\phi^2+\alpha^{-1}dr^2+(r^2+l^2)d\theta^2 \
,
\end{equation}
which can be received from the quadratic form~(\ref{gen_cylind})
with help of the following transformations
\bea \label{transform} &&f=\frac{N^2-(M^2+l^2)}{(N+M)^2+l^2}\ ,
\quad f^{-1}e^{2\gamma}=\frac{(N+M)^2+l^2}{p_{+}p_{-}}\ , \quad
r=N+M\ , \nonumber \\ &&N=\frac{1}{2}(p_{+}+p_{-})  \ ,  \quad
\cos\theta=\frac{p_{+}-p_{-}}{2A} \ , \qquad
\omega=\frac{l}{A}(p_{+}-p_{-})  \ , \nonumber \\
 &&\rho=\sqrt{(r-M)^2-A^2}\sin\theta \ , \quad z =
(r-M)\cos\theta \ ,
 \eea
used in paper~\cite{gauthoff}), parameter
$\alpha=(r^2-2Mr-l^2)/(r^2+l^2)$. Here parameters $M$ and $l$
represent gravitoelectric (total mass of the source) and
gravitomagnetic charges respectively. The parameter $l$ is also
called NUT parameter or the most commonly known as
'gravitomagnetic monopole'. From astrophysical point of view it is
reasonable to assume that the string has no electric charge and
consequently in the space-time around the nonspinning NUT string
the metric tensor has the following nonvanishing components
\bea &&g_{00}=-f\ , \qquad g_{11}=g_{33}=f^{-1}e^{2\gamma} \ ,
\nonumber
\\
&&g_{02}=f\omega \ , \qquad g_{22}=f^{-1}\rho^2-f\omega^2 \ . \eea

The Euler-Lagrange equations
\be \label{E-L} \frac{d}{d\lambda}\left(\frac{\pl {\cal L}}{\pl
\dot{x}^\alpha}\right) - \frac{\pl {\cal L}}{\pl x^\alpha}=0 \ ,
\ee
governing geodesics of the metric~(\ref{gen_cylind}) can be
derived from the Lagrangian
\be \label{lagr} 2{\cal
L}=g_{\mu\nu}\frac{dx^\mu}{d\lambda}\frac{dx^\nu}{d\lambda}\ . \ee
where a dot stands for differentiation with respect to an affine
parameter $\lambda$.

The integrals of motion
\bea \label{E} \frac{dp_t}{d\lambda}&=&\frac{\pl{\cal L}}{\pl
t}=0\ , \ p_t=-E\ , \nonumber \\ \nonumber\\ \label{L}
\frac{dp_\varphi}{d\lambda}&=&\frac{\pl{\cal L}}{\pl\varphi}=0\ ,
\ p_\varphi=L\  \eea
follow from Euler-Lagrange equations~(\ref{E-L}). Here
$p_\alpha=\pl {\cal L}/\pl \dot{x}^\alpha$ is canonical momenta
and the conserved quantities represent, respectively, the total
energy, orbital angular momentum of the test
particle~\cite{shapiro}.

Knowing that $g_{\alpha\beta}p^{\alpha}p^{\beta}=-m^2$ we can
conclude that value of $\cal L$ in~(\ref{lagr}) equals to
$-m^2/2$. For spacetime~(\ref{gen_cylind}) the
Lagrangian~(\ref{lagr}) is
\be \label{lagr_nut} {\cal L} = -\frac{1}{2}f\dot{t}^2
+f\omega\dot{\varphi}\dot{t}
    +\frac{1}{2}f^{-1}e^{2\gamma}({\dot{\rho}}^2 + {\dot{z}}^2)
    +\frac{1}{2}(f^{-1}\rho^2-f\omega^2){\dot{\varphi}}^2 \ ,
\ee
and the corresponding canonical momenta~(\ref{E}--\ref{lagr_nut})
are
\be \label{momenta} E = f({\dot{t}} - \omega{\dot{\varphi}})\ , \
L = f\omega{\dot{t}} + (f^{-1}\rho^2-f\omega^2){\dot{\varphi}} \ .
\ee

Our task here is complicated due to the fact that the metric
coefficients depend not only on radial coordinate, but also on
coordinate $z$ (the spacetime is axially symmetric). We have to
express $z$ through $\rho$.
 In the paper~\cite{bini} the motion of a free uncharged test
particle in the Kerr-Newman-Taub-NUT spacetime was considered.
Solving the Hamilton-Jacobi equation which describes the motion of
test particle in the vicinity of non rotating source endowed with
gravitomagnetic charge it was shown that the orbit will lie on a
cone with
\be \label{cos} \cos\theta=\pm \left| \frac{2lE}{L}\right|
 \ . \ee
As $l\rightarrow 0$ such an orbit degenerates to a single orbit on
the equatorial plane. Using transformations~(\ref{transform}) one
can get the expression for $z$ in cylindrical coordinates
\be \label{z} z=\frac{2lE}{L}
\sqrt{\frac{L^2(A^2+\rho^2)-4l^2A^2E^2 }{L^2-4l^2E^2}}
 \ . \ee

It is easy to see that $z=0$ corresponds to the equatorial plane
$\theta=\pi/2$ in spherical coordinates. In order to investigate
the radial dependence of effective potential  we restrict
ourselves to the study of circular geodesics neglecting possible
motion of particles along the $z$ axis, hence $\dot{z}=0$. This
implies that the equation~(\ref{lagr_nut}) is
\be
\label{geod_circ} -f\dot{t}^2 +2f\omega\dot{\varphi}\dot{t}
    +f^{-1}e^{2\gamma}{\dot{\rho}}^2
    +(f^{-1}\rho^2-f\omega^2){\dot{\varphi}}^2=-m^2 \ .
\ee
From equations~(\ref{momenta}) one could have
\be
\dot{t}=\frac{E(\rho^2-\omega^2f^2)+L\omega f^2}{f\rho^2} \ , \
\dot{\varphi}=\frac{L-E\omega}{f^{-1}\rho^2} \ . \ee

Consider a case when particle has a mass $m\neq0$. Then it is
convenient to normalize quantities $E$, $L$ to the unit of mass of
the particle $m$ and label them with a bar as $ \overline{E}
={E}/{m}$, $\overline{L}={L}/{m}$. The 'tilded' quantities are
normalized to the total mass of the source $M$. Then the equation
of circular motion~(\ref{geod_circ}) can be rewritten in the form
\be \label{R} f^{-1}e^{2\gamma}\left(\frac{1}{m}\frac{d \rho}{d
\lambda}\right)^2 =V(\rho,z) \ , \quad
V(\rho,z)\equiv-1
+\frac{\overline{E}^2}{f}-\frac{f(\overline{L}-\overline{E}\omega)^2}{\rho^2}\
. \ee

The elliptic integrals resulting from the solution of the
equations of motion are not particularly informative, but one can
get a general picture of the orbits by considering behaviour of an
effective potential. Define an effective potential $V_{eff}$ as
that value of $\overline{E}$ such that $\dot{\rho}=0$ at radius
$\rho$:
\be \label{V_eff} -1 +\frac{V_{eff}^2}{f}-\frac{f(\overline{L}/M
-V_{eff}\widetilde{\omega})^2}{\widetilde{\rho}^2}=0\ . \ee

The solution for effective potential
\be \label{V_eff_pot} V_{eff}=\frac{-\widetilde{\omega}
f^2(\overline{L}/M)\pm\sqrt{-
f(f^2\widetilde{\omega}^2-\widetilde{\rho}^2)\widetilde{\rho}^2+
f^2(\overline{L}/M)^2\widetilde{\rho}^2}}
    {\widetilde{\rho}^2-f^2\widetilde{\omega}^2} \
\ee
should have a sense of energy and a root which corresponds to the
asymptotic condition $\lim_{\rho \rightarrow \infty} V_{{eff}}=1$
has to be chosen.

Figure $1$ illustrates the dependence of the effective potential
on radial coordinate $\rho/M$ for a source endowed with mass $M$
and gravitomagnetic mass $l$.

\begin{figure}[hbtp] \label{pot_NUT}
\includegraphics[height=39mm, width=114mm]{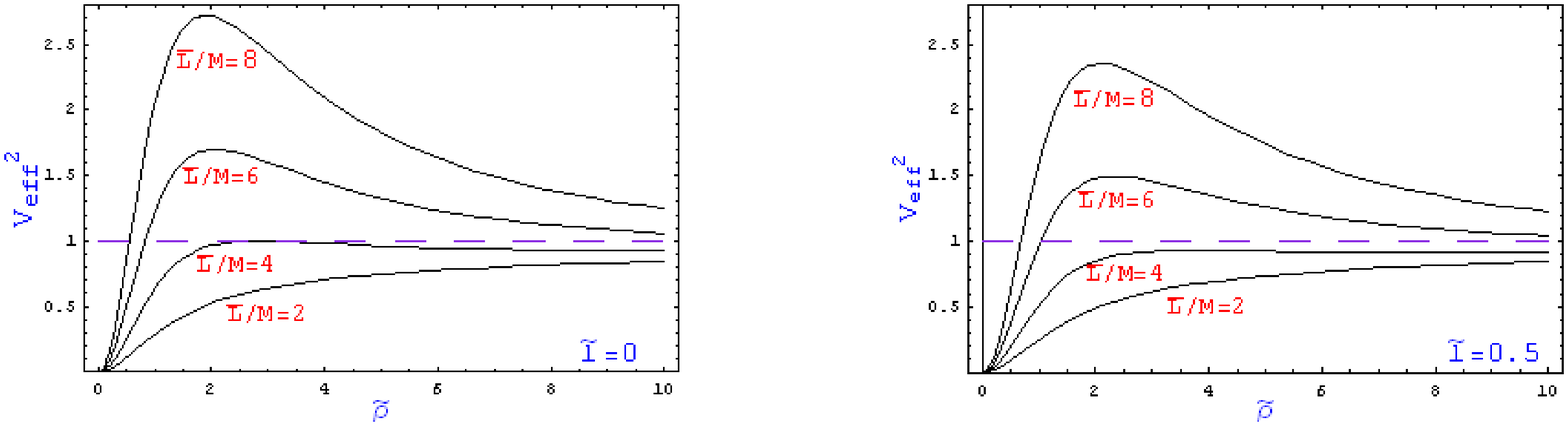} \caption{The radial
dependence of the effective potential for particles with nonzero
rest mass for different values of angular momentum $\overline{L}$.
The left hand side figure is responsible for the case when the NUT
parameter $\widetilde{l}=0$. For the right hand side figure the
NUT parameter $\widetilde{l}=0.5$. Maxima in the effective
potential indicate unstable circular orbits and minima stable
circular orbits. Curves for particles with equal angular momentum
and different gravitomagnetic charges have more monotonous
behavior with the increase of the value of NUT parameter.}
\end{figure}

To simplify further calculations we can make the assumptions
concerning the value of $\widetilde{l}$. We select second order
approximation in gravitomagnetic monopole moment which can be
justified by the fact that at the moment there is no any
astrophysical evidence for strong gravitomagnetic mass. For
example, in the recent papers~\cite{rahvar,rahnour} the magnetic
mass detection using the gravitational microlensing technique has
been explored and it has been evaluated that the minimum
observable gravitomagnetic mass to be about $14 meters$. The
following assumption is made concerning energy $\overline{E}$ and
angular momentum $\overline{L}$ of the particle. As the quantity
$\widetilde{l}^2$ is small then for the nonrelativistic particles
with small values of energy and angular momentum one can neglect
terms being proportional to $\overline{E}^2\widetilde{l}^2$ and
$\overline{L}^2\widetilde{l}^2$ and get
\be \label{z_appr}
\widetilde{z}=\frac{2\widetilde{l}\overline{E}}{\overline{L}/M}
\sqrt{1+\widetilde{\rho}^2} \ . \ee

 The conditions for circular motion of test particle at radius
$\rho=\rho_0$ are
 \be
\label{stabil}
 V(\rho_0)=0\ , \ {V}^\prime|_{\rho=\rho_0}=0\ ,
\ee
where prime denotes the derivative on radial coordinate.
Furthermore, the circular motion is stable if
\be {V}''|_{\rho=\rho_0}<0 \ . \ee
From the equation~(\ref{stabil}) after some lengthly algebra one
can obtain the following expression for the energy of the particle
 \be
 \label{energy_NUT}
\overline{E}^2\approx
\frac{2\widetilde{\rho}^2(X+\widetilde{\rho}^2)
\left[2\widetilde{l}^2\left(X^2+\widetilde{\rho}^2(1+\widetilde{\rho}^2)\right)+
X\widetilde{\rho}^2(X+\widetilde{\rho}^2)\right]}{X^3
\left[2(1+\widetilde{\rho}^2)\left(\widetilde{\rho}^4+\widetilde{\rho}^2(X-2)-4X\right)
+\widetilde{l}^2Y\right]} \ , \ee
where $X=1+\sqrt{1+\widetilde{\rho}^2}$ and
$Y=3\widetilde{\rho}^4(X-2)+10X+6\widetilde{\rho}^2X.$ The
circular motion is possible provided that the expression for
$\overline{E}^2$ is nonnegative~\cite{aliev}. Circular orbits
exist till a limiting case corresponding to the photon orbit
($\overline{E}=E/m\rightarrow \infty$). When NUT parameter is
equal to zero this limit is equal to
$\rho^2=\widetilde{\rho}^2M^2=3M^2$. This corresponds to the value
$r=3M$ in spherical symmetric Schwarzschild spacetime. The
influence of NUT parameter is such that the limit of existence of
circular orbits decreases. For example, for $\widetilde{l}=0.5$ we
have $\widetilde{\rho}^2=2.49$.

The orbit is maximally bound when $\overline{E}^2=1$. Equating the
expression~(\ref{energy_NUT}) to unity one can obtain the equation
governing the bounds of maximally binding orbits. In the field of
a cylindrical source endowed with NUT parameter, taken as
$\widetilde{l}=0.1$, radius of maximally binding orbit is
$\widetilde{\rho}^2\approx 7.98$. Under the influence of NUT
parameter the value of maximal radius of bound orbits decreases.
For comparison: when $\widetilde{l}=0$ radius of the orbit is
$\widetilde{\rho}^2=8$ which corresponds to $r=4M$ in
Schwarzschild spacetime.

The orbit situated at an inflection point of the effective radial
potential, that is, with
\be
\label{sec_deriv_R} {V}''|_{\rho=\rho_0}=0  \
\ee
  is the least tightly bound and
describes the bounds of stability region.The solution of this
equation for $\widetilde{l}=0$ is $\widetilde{\rho}^2=24$. This
corresponds to $r=6M$ for spherical symmetric Schwarzschild case.
The orbits are stable for $\widetilde{\rho}>2\sqrt{6}$ and
unstable for $\widetilde{\rho}<2\sqrt{6}$. For $\widetilde{l}=0.7$
we have for limit of stability  $\widetilde{\rho}^2=34.41$.
Therefore, the radius of critical orbit separating stable orbits
from unstable for nonzero gravitomagnetic monopole increases.

The influence of the NUT parameter on the motion of a test
particle can be seen from the following table
\begin{center}
\begin{tabular}{|l|l|l|l|} \hline\hline
\multicolumn{4}{|c|}{Gravitomagnetic influence on motion} \\
\hline $\widetilde{l}$ & $\widetilde{\rho}^2_{circular}$ &
$\widetilde{\rho}^2_{max.bound}$
 & $\widetilde{\rho}^2_{stable}$  \\
\hline 0 & 3 & 8        & 24 \\ \hline 0.05 & 2.99        & 7.99
& 24.02 \\ \hline 0.1 & 2.98        & 7.98        & 24.08 \\
\hline 0.5 & 2.49 & 7.5        & 27.21 \\ \hline 0.7 & 1.97 &
6.94& 34.41
\\ \hline\hline
\end{tabular}
\end{center}
where $\widetilde{\rho}^2_{circular}$ defines the radius of last
circular orbit, $\widetilde{\rho}^2_{max.bound}$ defines the
radius of the last bound orbit and $\widetilde{\rho}^2_{stable}$
defines the radius of first stable orbit.

Hereafter for simplicity we will omit widetilded and overlined
labels but keep in mind that all quantities are normalized to the
mass of the source, energy and angular momentum are also
normalized to the mass of the particle. The four-velocity of the
proper observer has the following nonvanishing components
 \begin{equation}
u^\alpha \equiv \frac{1}{\sqrt{f}} \Bigg\{1\ , 0\ , 0\ , 0
\Bigg\}\ , \qquad
  u_\alpha \equiv \sqrt{f}
\Bigg\{-1\ , 0\ , \omega \ , 0\Bigg\} \ .
\end{equation}

Components of absolute acceleration $w_\alpha =u_{\alpha
;\beta}u^\beta$ of the proper observer in the gravitational field
~(\ref{gen_cylind}) are given by
\begin{equation}
\label{accel} w_\alpha\equiv \frac{1}{2f} \Bigg\{0\ , f_{,\rho}\
,0\ ,f_{,z}\Bigg\}\ .
\end{equation}

 The nonvanishing components of the relativistic rate of
rotation $A_{\beta\alpha}=u_{[\alpha
,\beta]}+u_{[\beta}w_{\alpha]}$ are
\begin{equation}
\label{A} A_{r\varphi}= \frac{1}{2}\sqrt{f}\omega_{,\rho}
 \ ,
\qquad A_{z\varphi}= \frac{1}{2}\sqrt{f}\omega_{,z}\ .
\end{equation}

For the case of zero gravitomagnetic charge all components of
relativistic rate of rotation are equal to zero. It means that NUT
parameter influences on the observer as a force which drags the
observer and makes him rotate around the source.

The geodesic equations $\delta^2x^\alpha/\delta\sigma^2=0$ in
spacetime~(\ref{gen_cylind}) take the form
\bea
 \label{geod_0}
&&\ddot{t}+\frac{\dot{t}}{f\rho^2}
   \left[\rho^2\mu_f
    +f^3\omega\mu_{\omega}\right]
    \nonumber\\
    &&+\frac{\dot{\varphi}}{f\rho^2}
    \left[2f\omega\rho\dot{\rho}-2\rho^2\omega\mu_f
    -f(\rho^2+f^2\omega^2)\mu_{\omega}\right]=0 \ ,
\\ \nonumber\\
 \label{geod_1}
&&\ddot{\rho}+(\gamma_{,\rho}-\frac{f_{,\rho}}{2f})(\dot{\rho}^2-\dot{z}^2)
     +\frac{e^{-2\gamma}}{2f}\left[f^2f_{,\rho}\dot{t}^2+(-2\rho f
     +\xi_{\rho})\dot{\varphi}^2\right]
\nonumber\\
     &&+2(\gamma_{,z}-\frac{f_{,z}}{2f})\dot{\rho}\dot{z}
     -e^{-2\gamma}f(\omega f_{,\rho}+f\omega_{,\rho})\dot{\varphi}\dot{t}=0 \ ,
\\ \nonumber\\
\label{geod_2}
&&\ddot{z}+(\gamma_{,z}-\frac{f_{,z}}{2f})(\dot{z}^2-\dot{\rho}^2)
     +\frac{e^{-2\gamma}}{2f}
     \left[\xi_{z}\dot{\varphi}^2
     +f^2f_{,z}\dot{t}^2\right]
     \nonumber\\
     &&-e^{-2\gamma}f(\omega f_{,z}
     +f\omega_{,z})\dot{\varphi}\dot{t}
     +2(\gamma_{,\rho}-\frac{f_{,\rho}}{2f})\dot{\rho}\dot{z} =0
\ ,
\\ \nonumber\\
\label{geod_3} &&\ddot{\varphi}+\frac{\dot{t}f^2}{\rho^2}
   \mu_{\omega}
   +\frac{\dot{\varphi}}{f\rho^2}(2f\rho\dot{\rho}
    -\rho^2(f_{,\rho}\dot{\rho}+f_{,z}\dot{z})
    -f^3\omega\mu_{\omega})=0 \ ,
    \eea
    where $\delta$ denotes the covariant derivative and $\mu_f=
f_{,\rho}\dot{\rho}+f_{,z}\dot{z}$,
$\mu_{\omega}=\omega_{,\rho}\dot{\rho}+\omega_{,z}\dot{z}$,
$\xi_{\rho}=(\rho^2+f^2\omega^2)f_{,\rho}+2f^3\omega
     \omega_{,\rho}$, $\xi_{z}=(\rho^2+f^2\omega^2)f_{,z}+2f^3\omega
     \omega_{,z}\ .$

 For the circular geodesics $\dot{\rho}=\dot{z}=0$ only radial ~(\ref{geod_1}) and $z$
component~(\ref{geod_3}) of the geodesic equations are not
vanishing
\bea \label{geod_1_circ} &&f_{,\rho}\dot{t}^2+(-\frac{2\rho}{f}
     +f^{-2}\xi_{\rho})\dot{\varphi}^2
    -2(\omega
    f_{,\rho}+f\omega_{,\rho})\dot{\varphi}\dot{t}=0 \ ,
\\ \nonumber\\
\label{geod_2_circ} &&f^{-2}\xi_{z}\dot{\varphi}^2
     +f_{,z}\dot{t}^2-2(\omega f_{,z}
     +f\omega_{,z})\dot{\varphi}\dot{t}=0
\ . \eea

The angular velocity of the test particle is given by
$\varpi=d\varphi/dt=d\varphi/d\sigma\left(dt/d\sigma\right)^{-1}=
\dot{\varphi}/\dot{t}$ which becomes, using~(\ref{geod_1_circ})
and ~(\ref{geod_2_circ}) as
\bea  \label{varpi1}
     \varpi&=&f\frac{\kappa \pm\sqrt{\kappa^2
     -(f_{, \rho} + f_{, z})\chi}}{\chi} \ , \eea
where
\begin{equation}
\kappa=f[f(\omega_{, \rho} + \omega_{, z}) + \omega(f_{, \rho} +
f_{, z})]
\end{equation}
 and
\begin{equation}
 \chi=(f_{, \rho} + f_{, z})(\rho^2+f^2\omega^2)
     -2f\rho+2f^3\omega(\omega_{, \rho} + \omega_{, z})\ .
\end{equation}
Equation~(\ref{varpi1}) defines the angular velocity of a zero
angular momentum particle. Due to the effect of dragging of
inertial frames the particles are dragged by the influence of NUT
parameter. This effect weakens with distance as $1/\rho^3$ and it
makes the gravitomagnetic charge of source measurable in
principle.

\section{Electromagnetic Fields of \\ Nonrotating Cylindrical NUT Source}
\label{meq}

   The general form of the first pair of general
relativistic Maxwell equations is given by
\begin{equation}
\label{maxwell_firstpair}
F_{\alpha \beta, \gamma }
    + F_{\gamma \alpha, \beta} + F_{\beta \gamma,\alpha}
     = 0 \ ,
\end{equation}
where $F_{\alpha \beta}$ is the electromagnetic field tensor
expressing the strict connection between the electric and magnetic
four-vector fields $E^{\alpha},\ B^{\alpha}$. For an observer with
four-velocity $u^{\alpha}_{(obs)}$, the covariant components of
the electromagnetic tensor are given by
\begin{equation}
\label{fab_def} F_{\alpha\beta} \equiv 2 u_{(obs)[\alpha}
E_{\beta]} +
    \eta_{\alpha\beta\gamma\delta}u_{(obs)}^\gamma B^\delta \ ,
\end{equation}
where $\eta_{\alpha\beta\gamma\delta}$ is the pseudo-tensorial
expression for the Levi-Civita symbol $\epsilon_{\alpha \beta
\gamma \delta}$
\begin{equation}
\label{eta} \eta^{\alpha\beta\gamma\delta}=-\frac{1}{\sqrt{-g}}
    \epsilon_{\alpha\beta\gamma\delta} \ ,
    \hskip 2.0cm
\eta_{\alpha\beta\gamma\delta}=
    \sqrt{-g}\epsilon_{\alpha\beta\gamma\delta} \ ,
\end{equation}
with $g\equiv {\rm det}|g_{\alpha\beta}|= -{e^{4\gamma}}\rho^2
f^{-2}$ for the metric (\ref{gen_cylind}).

A useful class of observers is represented by the ``zero angular
momentum observers'' or ZAMOs~\cite{bpt72}. These are observers
that are locally stationary (i.e. at fixed values of $r$ and
$\theta$) but who are ``dragged'' into rotation with respect to a
reference frame fixed with respect to distant observers. They have
four-velocity components given by
\begin{equation}
\label{uzamos} (u^{\alpha})_{_{\rm ZAMO}}\equiv
   b^{-1}\bigg(1,0,-\frac{f^2\omega}{a},
    0 \bigg) \ ;
    \hskip 0.5cm
(u_{\alpha})_{_{\rm ZAMO}}\equiv
    b\bigg(- 1,0,0,0 \bigg) \ ,
\end{equation}
where $a=\rho^2-f^2\omega^2$, $b=\rho\sqrt{f/a}$.

The general form of the second pair of Maxwell equations is given
by
\begin{equation}
\label{maxwell_secondpair} F^{\alpha \beta}_{\ \ \ \ ;\beta} =
4\pi J^{\alpha}\ ,
\end{equation}
where $J^{\alpha}$ is the four-current.

Maxwell equations assume a familiar flat-spacetime form when
projected onto a locally orthonormal tetrad. In principle such
tetrad is arbitrary, but in the case of a relativistic rotating
metric source a ``natural'' choice is offered by the tetrad
carried by the ZAMOs. Using (\ref{uzamos}) we find that the
components of the tetrad $\{{\bf e}_{\hat \mu}\} = ({\bf e}_{\hat
0},{\bf e}_{\hat \rho},{\bf e}_{\hat \phi}, {\bf e}_{\hat z})$
carried by a ZAMO observer are
\begin{eqnarray}
\label{zamo_tetrad_0} &&{\bf e}_{\hat0}^{\alpha} =
     b^{-1}\bigg(1,0,-\frac{f^2\omega}{a},0\bigg) \ ,      \\
\label{zamo_tetrad_1} &&{\bf e}_{\hat \rho}^{\alpha}  =
    \sqrt{f}e^{-\gamma}\bigg(0,1,0,0\bigg) \ ,        \\
\label{zamo_tetrad_2} &&{\bf e}_{\hat \phi}^{\alpha}  =
   \sqrt{\frac{f}{a}}\bigg(0,0,1,0\bigg)  \ ,         \\
\label{zamo_tetrad_3} &&{\bf e}_{\hat z}^{\alpha}  =
   \sqrt{f}e^{-\gamma}\bigg(0,0,0,1\bigg) \ .
\end{eqnarray}

    We can now rewrite the Maxwell equations
 (\ref{maxwell_firstpair}) and (\ref{maxwell_secondpair})
 in the ZAMO reference
frame by contracting  them
 with
(\ref{zamo_tetrad_0})--(\ref{zamo_tetrad_3}). After some lengthy
but straightforward algebra, we obtain Maxwell equations in the
more useful form
\begin{eqnarray}
\label{max1a} &&\frac{e^{2\gamma}}{f} B^{\hat \phi}_{,\phi}+
\left(\frac{e^{\gamma}\sqrt{a}}{f}B^{\hat z}\right)_{, z}+
\left(\frac{e^{\gamma}\sqrt{a}}{f}B^{\hat \rho} \right)_{, \rho} =
0 \ ,
\\
\nonumber\\ \nonumber\\
 \label{max1b} &&\frac{\partial B^{\hat
\rho}}{\partial t}
    =\frac{f}{a}\left(\rho E^{\hat z}_{,\phi}
    +\omega fB^{\hat \rho}_{\
,\phi}\right)-\frac{e^{-\gamma}f\rho}{\sqrt{a}} E^{\hat \phi}_{,z}
\ ,
\\
\nonumber\\ \nonumber\\
 \label{max1c}
    &&\frac{\partial B^{\hat \phi}}{\partial t}
    = \frac{f}{e^{2\gamma}}\left[\rho\left(\frac{e^{\gamma}}{\sqrt{a}}E^{\hat \rho}
    -\frac{e^{\gamma}\omega f}{\sqrt{a}}B^{\hat z}\right)_{
,z}-\left(\frac{e^{\gamma}\rho}{\sqrt{a}}E^{\hat z}
-\frac{e^{\gamma}\omega f}{\sqrt{a}}B^{\hat \rho}\right)_{
,\rho}\right] \ ,
\\
\nonumber\\ \nonumber\\ \label{max1d}
    &&\frac{\partial B^{\hat z}}{\partial t}
    =\frac{e^{-\gamma}f}{\sqrt{a}}\left(\rho E^{\hat \phi}\right)_{\ ,\rho}-
   \frac{f}{a}\left(\rho E^{\hat \rho}_{,\phi}
    -\omega fB^{\hat z}_{\
,\phi}\right) \ ,
\end{eqnarray}
\noindent
and
\begin{eqnarray}
\label{max2a} &&\left(e^{\gamma}\frac{\sqrt{a}}{f} E^{\hat \rho}
\right)_{,\rho}+
    \frac{e^{2\gamma}}{f}E^{\hat \phi}_{,\phi}+\left(e^{\gamma}\frac{\sqrt{a}}{f} E^{\hat z}
\right)_{, z}
    =  4\pi \frac{e^{2\gamma}}{f^{3/2}}\sqrt{a}J^{\hat t} \ , \\
\nonumber\\ \nonumber\\ \label{max2b}
&&\frac{e^{\gamma}}{\sqrt{a}}\left(
    \omega fE^{\hat \rho}_{\
,\phi}-\rho B^{\hat z}_{,\phi}\right)+\rho B^{\hat \phi}_{,
z}=\frac{e^{\gamma}\sqrt{a}}{f}\frac{\partial E^{\hat
\rho}}{\partial t}+4\pi \frac{e^{\gamma}\rho}{\sqrt{f}}J^{\hat
\rho} \ ,
\\
\nonumber\\ \nonumber\\ \label{max2c} &&
-\left[\frac{e^{\gamma}}{\sqrt{a}}(
    \omega fE^{\hat \rho}-\rho B^{\hat z})
    \right]_{, \
\rho}-\left[\frac{e^{\gamma}}{\sqrt{a}}(
    \omega fE^{\hat z}+\rho B^{\hat \rho})\right]_{, \ z}
 \nonumber \\ \nonumber\\
 &&= \frac{e^{2\gamma}}{f}
    \frac{\partial E^{\hat \phi}}{\partial t}
    +4\pi \frac{e^{2\gamma}}{\sqrt{f a}}(\rho J^{\hat \phi}-f \omega J^{\hat t}) \ ,
   \\
\nonumber\\ \nonumber\\
    \label{max2d}
    &&\frac{e^{\gamma}}{\sqrt{a}}\left(
    \omega fE^{\hat z}_{, \phi}+\rho B^{\hat \rho}_{, \phi}\right)- (\rho B^{\hat
\phi})_{,\rho}=\frac{e^{\gamma}\sqrt{a}}{f}
    \frac{\partial E^{\hat z}}{\partial t}
+ 4\pi \frac{e^{\gamma}\rho}{\sqrt{f}} J^{\hat z} \ .
\end{eqnarray}
%


We will look for stationary and axially symmetric solutions of the
Maxwell equations, taking into account that in the vacuum region
around the source all components of electric current are equal to
zero. We could study the realistic configuration of magnetic
fields, for example, the dipolar magnetic field. Due to the
possible smallness of the gravitomagnetic mass of the star
estimated from some astrophysical observations we may perform
calculations to the first order in NUT parameter.

%
%

\begin{figure}[hbtp] \label{pot_NUT}
\includegraphics[height=76mm, width=114mm]{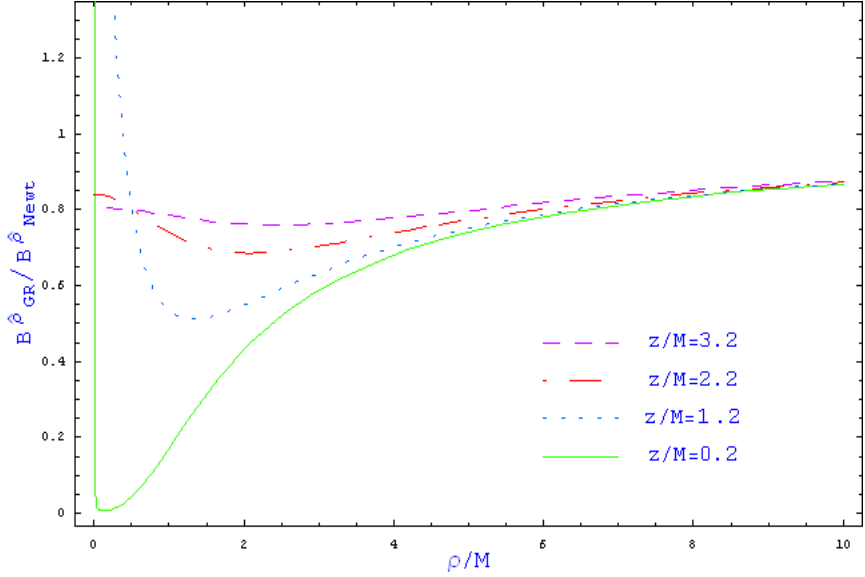} \label{magnetic_field_NUT}
\caption{Dependence of general relativistic modification factor $B^{\hat
\rho}_{GR}/B^{\hat \rho}_{Newt}$ of magnetic field on the radial
coordinate $\rho/M$ normalized in units of stellar mass. Near to
the NUT source the magnetic field will be amplified, then in some
intermediate region it will be weakened and in the asymptotically
far zone the influence of NUT parameter is negligible, the
behavior of field is Newtonian and relation tends to unity. The
influence of the NUT parameter is more strong near to the source
of the $z$ axis.}
\end{figure}

According to the transformations~(\ref{transform}) solution for
vector potential in spherical coordinates (see, for example,
~\cite{wasserman,rezzolla_ahmedov})
\begin{equation}
  A_{\phi}(r, \theta)
= -\frac{3\mu \sin^2\theta}{2M}
 \left[\left(\frac{r}{2M}\right)^2\ln(1-\frac{2M}{r})
 +\frac{1}{2}\left(\frac{r}{M}+1\right)\right]
\end{equation}
can be rewritten in cylindrical ones as
\begin{equation}
 A_{\phi}(\rho, z) = -\frac{3\mu}{8M^3}\frac{\rho^2}{f}
 \left[\ln f+\frac{2M(2M+N)}{(M+N)^2}\right] \ ,
\end{equation}
where $\mu$ is the dipolar magnetic moment.

This yields the following
%
exact solutions for the nonvanishing components of magnetic field
\bea \label{B_rho}
 B^{\hat \rho}&=&-\frac{3\mu\rho e^{-\gamma}}{8M^3f(M+N)^3}
\Big\{2MfN_{, z}(N+3M) \nonumber \\ &+&f_{,
z}(M+N)\left[(M+N)^2\ln f+3M^2-N^2\right]\Big\}\ ,
\\
\nonumber\\
\label{B_z}
 B^{\hat z}&=&\frac{3\mu e^{-\gamma}}{8M^3f(M+N)^3}
\Big\{ 2Mf\rho N_{, \rho}(N+3M) \nonumber \\
&+&(M+N)\Big[2M(N+2M)(\rho f_{,\rho}-2f) \nonumber \\
&+&(M+N)^2\left[\rho f_{,\rho}(\ln f-1)-2f\ln f\right]\Big]\Big\}
 \ .
\eea

\begin{figure}[hbtp] \label{pot_NUT}
\includegraphics[height=39mm, width=114mm]{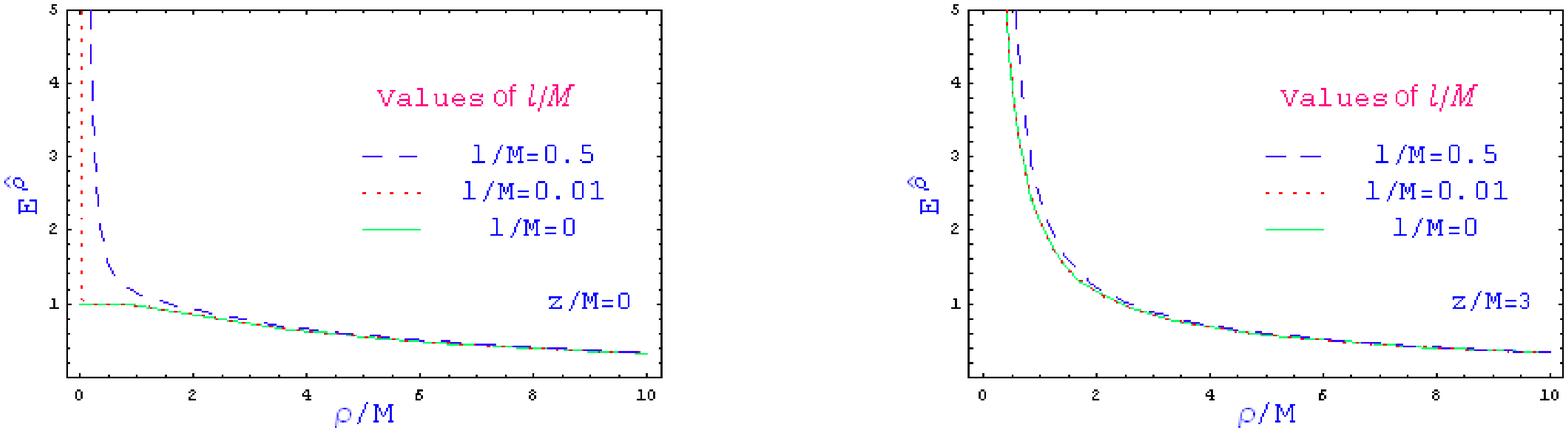} \label{electric_field_NUT}
\caption{The radial dependence of electric field $E^{\hat \rho}$
for different values of the gravitomagnetic monopole $l$. The
effect of the NUT parameter on the electric field is becoming
important near to the source of the $z$ axis.}
\end{figure}

The dependence of relation $B^{\hat \rho}_{GR}/B^{\hat
\rho}_{Newt}$ on parameter $\rho/M$ is shown on
Figure~\ref{magnetic_field_NUT}.

In the limit of flat spacetime when $\rho=r\sin\theta$,
$z=r\cos\theta$ and $\lim_{M \rightarrow 0} =0$ the
expressions~(\ref{B_rho})-(\ref{B_z}) reduce to the Newtonian
ordinary expressions for magnetic field of a magnetic
dipole~\cite{ll}
\be B^{\hat \rho}=\frac{\mu(2z^2-\rho^2)}{(z^2+\rho^2)^{5/2}} \ ,
\qquad B^{\hat z}=\frac{3\mu z\rho}{(z^2+\rho^2)^{5/2}} \ .
 \ee

Consider now  as a toy model the charged NUT star with the
monopolar electric field configuration
\bea &&E^{\hat \rho}=E^{\hat \rho}(\rho,z)\neq 0 \ , \\ &&E^{\hat
z}=E^{\hat \phi}=0 . \
 \eea
If $Q$ is the  electric charge per unit length of the line tube
then the solution for radial electric field admitted by Maxwell
equations is
\be
E^{\hat \rho}=Q\frac{fe^{-\gamma}}{\sqrt{a}} \ . \ee
Figure~\ref{electric_field_NUT} shows that for small values of $z$
and $\rho$ (near to the source) the influence of NUT parameter is
noticeable.

\section{Conclusion}
\label{conclusion}

We have investigated the general relativistic effects associated
with the possible existence of the gravitomagnetic monopole moment
of cylindrical NUT source and in particular studied the influence
of the gravitomagnetic monopole moment on the motion of the test
particles along circular orbits. These effects could be divided
into two parts. First effect is related to the fact that due to
the influence of the NUT parameter the particle's circular orbits
will be not kept in the equatorial plane. Our aim was mainly
related to the investigation of the second effect connected with
the influence of the NUT parameter on the circular orbits of the
particles. It is shown that under the influence of the NUT
parameter the particle's orbits are becoming less stable which can
be seen from the table presented in the paper.

Analytical general relativistic expressions for the dipolar
magnetic field around NUT star are presented. It is shown that in
the linear approximation in NUT parameter there is only effect of
mass of the star on the value of stationary magnetic field. In the
case of charged star there is strong influence of NUT parameter on
the value of monopolar electric field. Our results in principle
could be combined with analysis of astrophysical data on
electromagnetic fields of compact objects in order to get upper
limits on the value of NUT parameter.

\section*{Acknowledgments}
BA wishes to thank Naresh Dadhich for useful discussions. VK
gratefully acknowledges Claus L\"{a}mmerzahl and Jutta Kunz for
helpful comments and DFG for the financial support towards her
stay in ZARM, Bremen, and Oldenburg University.
 Financial support for this work is
partially provided by the Abdus Salam International Centre for
Theoretical Physics through associateship program and grant AC-83.
This research is also supported by the NATO Reintegration Grant
EAP.RIG.981259, by Uz FFR (project 1-06) and projects F2.1.09 and
F2.2.06 of the Uz CST.

\newpage

\end{document}